\documentclass[aps,prl,reprint]{revtex4-1}

\pdfoutput=1

\usepackage{mathtools}
\usepackage{amssymb}
\usepackage{xcolor}
\usepackage[colorlinks, citecolor={blue!70!black}, urlcolor={blue!70!black}, linkcolor={red!70!black},hyperindex,breaklinks]{hyperref}

\newcommand{\beq}{\begin{equation}}
\newcommand{\eeq}{\end{equation}}
\renewcommand{\hbar}{\mathchar'26\mkern-9mu h}
\newcommand{\w}{\omega}

\newcommand{\Sp}{S' \hspace{-0.1em}}
\newcommand{\Spp}{S'' \hspace{-0.1em}}
\DeclareMathOperator{\Tr}{Tr}

\begin{document}

\title{Bounds on chaos from the eigenstate thermalization hypothesis}

\author{Chaitanya Murthy}
\email{cm@physics.ucsb.edu}
\author{Mark Srednicki}
\email{mark@physics.ucsb.edu}
\affiliation{Department of Physics, University of California, Santa Barbara, CA 93106}

\date{\today}

\begin{abstract}
We show that the known bound on the growth rate of 
the out-of-time-order four-point correlator in chaotic many-body quantum systems 
follows directly from the general structure of operator matrix elements in systems 
that obey the eigenstate thermalization hypothesis.
This ties together two key paradigms of thermal behavior
in isolated many-body quantum systems.
\end{abstract}

\maketitle

In recent years there has been renewed interest in various ways of quantifying the rates of runaway growth processes
in quantum chaotic many-body systems. One particular quantity that has been studied extensively is the
four-point out-of-time-order (OTO) correlator \cite{Larkin1969,Kitaev2015,Maldacena2016};
for a recent overview, see Ref.~\cite{Swingle2018}. 
Here, following Ref.~\cite{Maldacena2016}, we consider the thermally regulated OTO correlator,
\beq
F_{\textsc{oto}}(t) \coloneqq \Tr\bigl[\rho^{1/4}\! A(t)\rho^{1/4}\kern-2pt A(0)\rho^{1/4}\kern-2pt A(t)\rho^{1/4}\kern-2pt A(0)\bigr] ,
\label{eq:FO}
\eeq
where $A(t)=e^{iHt}\!A e^{-iHt}$ is a local operator in the Heisenberg picture,
$H$ is the hamiltonian, 
$\rho \coloneqq e^{-\beta H}\!/Z$ is a thermal density operator at inverse temperature $\beta$, and 
$Z\coloneqq\Tr e^{-\beta H}$ is the partition function. (We set $\hbar = k_B = 1$ throughout.)
Following Refs.~\cite{Kitaev2015,Maldacena2016}, we consider systems with a scrambling time (also called the Ehrenfest time)
$t_\mathrm{s}$ that is large compared to the dissipation time $t_\mathrm{d}$ that governs the exponential decay rate of the two-point correlator.
In this case, for times $t_\mathrm{d} \ll t \ll t_\mathrm{s}$, we expect
\beq
F_{\textsc{oto}}(t) \propto 1 - e^{\lambda (t-t_\mathrm{s})},
\label{eq:FO2}
\eeq
where $\lambda$ is a growth rate that is analogous to the Lyapunov growth rate of 
the deviation of nearby classical trajectories in chaotic systems.
The hierarchy of time scales $t_\mathrm{s} \gg t_\mathrm{d}$ typically arises only in
systems that have a small parameter $\epsilon$ that determines $t_\mathrm{s}$
via $t_\mathrm{s}\sim \lambda^{-1}\ln(1/\epsilon)$.
Examples include $\epsilon\sim 1/N^2$ for the Sachdev--Ye--Kitaev model
of $N\gg 1$ Majorana fermions with all-to-all random four-point interactions  \cite{Sachdev1992,Kitaev2015},
and for conformal field theories with $N^2\gg 1$ fields 
that have gravitational duals \cite{Maldacena2016},
and $\epsilon \sim \hbar_\mathrm{eff}$, an effective dimensionless Planck's constant,
for semiclassical systems such as the kicked rotor \cite{Rozenbaum2017}
and quantized area-preserving maps \cite{Garcia-Mata2018}.

Making a set of physical and mathematical assumptions that are plausible in such systems,
Maldecena \textit{et al.}~\cite{Maldacena2016} argued that the growth rate $\lambda$ should be bounded by
\beq
\lambda \le 2\pi /\beta.
\label{eq:lam}
\eeq
This bound is saturated in the SYK model, and in large-$N$ conformal field theories with gravitational duals,
where it is related to the physics of information scrambling in black holes \cite{Shenker2014}.

Another paradigm that is believed to apply broadly to many-body quantum systems 
with sufficiently strong interactions and no disorder (or, more generally, disorder that
does not result in many-body localization \cite{Nandkishore2015}) is the
\emph{eigenstate thermalization hypothesis} (ETH) \cite{Deutsch1991,Srednicki1994,Srednicki1999,Rigol2008,DAlessio2016},
which supposes that the energy eigenstates of such a system cannot be distinguished from a thermal density matrix 
when probed by local observables. 

More precisely, according to ETH, the matrix elements of a local observable $A$ in the energy-eigenstate basis,
$H|i\rangle = E_i|i\rangle$, 
take the form
\beq
A_{ij} = \mathcal{A}(E)\delta_{ij} + e^{-S(E)/2}f(E,\w)R_{ij},
\label{eq:Aij}
\eeq
where $E=(E_i+E_j)/2$ is the average energy of the two eigenstates, $\w=E_i-E_j$ is the energy difference,
${\cal A}(E)=\Tr\rho A$ with $\beta$ fixed by $E=\Tr\rho H$,
$S(E)$ is the thermodynamic entropy (logarithm of the density of states) at energy $E$,
$f(E,\w)$ is a smooth, real function of its two arguments
with $f(E,\w)=f(E,-\w)$, 
and $R_{ij}$ is a hermitian matrix of erratically varying elements, 
with overall zero mean and unit variance in local ranges of $E$ and $\w$.
It is consistent (as will be seen below) to treat $E$ as an extensive quantity
and $\w$ as an intensive quantity. 

Our purpose is to derive the bound on the OTO correlator growth rate, Eq.~(\ref{eq:lam}), 
directly from ETH. In doing so, we relate two important paradigms of 
thermal behavior in isolated many-body quantum systems.

Our methodology is to use known properties of the ETH matrix elements to put a bound on the
Fourier transform of the OTO correlator (more specifically, on its connected part, defined below) at high frequencies. 
This bound can then be used to infer bounds on the OTO correlator itself at intermediate times,
with some additional dependence on its precise functional form. For a family of functions that includes
the OTO correlator for a conformal field in one spatial dimension,
as computed by Maldacena \textit{et al.}~\cite{Maldacena2016b} via the $\mathrm{AdS}_2$ gravity dual, we find that
the bound of Eq.~(\ref{eq:lam}) must hold. For a simpler family of functions that is sometimes used
as an approximation to the OTO correlator, we find a stronger bound, indicating that this approximation should be
used with care. Our methods do not require the factorization assumption that was used in
Ref.~\cite{Maldacena2016} (and which we review below). 
Hence we believe that our result is more general, and that the bound on the exponential growth rate of the OTO correlator
holds in any quantum many-body system that obeys ETH
and also has the hierarchy of time scales $t_\mathrm{s}\gg t_\mathrm{d}$.

Note that we do not claim that ETH is either necessary or sufficient to have exponential growth of the OTO correlator;
we claim only that the exponential growth rate, if nonzero, is bounded by Eq.~(\ref{eq:lam}) in systems that obey ETH.
Also, it may also be possible to weaken our assumptions and still prove the bound; 
for example, our proof would go through if we allowed the envelope function $f(E,\w)$ to become noisy 
(rather than smooth) at sufficiently low frequencies $\w$. 

It will be most convenient to work with an observable $A$ for which
${\cal A}(E)=0$, either due to a symmetry, 
or simply by subtracting $\Tr\rho A$ from $A$; we therefore take ${\cal A}(E)=0$ from here on.

We begin by considering a thermally regulated two-point correlator for such an observable $A$ at inverse temperature $\beta$,
\beq
F_2(t) \coloneqq \Tr\bigl[\rho^{1/2}\kern-2pt A(t)\rho^{1/2} \kern-2pt A(0)\bigr] .
\label{eq:F2}
\eeq 
Inserting two complete sets of energy eigenstates and using Eq.~(\ref{eq:Aij})
with ${\cal A}(E)=0$,
we have
\beq
F_2(t) = \frac{1}{Z} \sum_{ij}e^{-S(E)-\beta E}|f(E,\w)|^2 e^{i\w t} |R_{ij}|^2. 
\label{eq:F2a}
\eeq
We replace $|R_{ij}|^2$ with its statistical average $1$, and then write each sum as an integral with a suitable density of states, 
$\sum_i \to \int_0^\infty dE_i\,e^{S(E_i)}$.
Using $E_{i,j} = E \pm \omega/2$, we get
\begin{align}
F_2(t) = \frac{1}{Z} \int_E \int_{\w} \ &e^{S(E+\w/2)+S(E-\w/2)-S(E)-\beta E} \nonumber \\*
&\times |f(E,\w)|^2 e^{i\w t},
\label{eq:F2b}
\end{align}
where $\int_E \coloneqq \int_0^{\infty} dE$ and $\int_{\w} \coloneqq \int_{-\infty}^{+\infty} d\w$.
We now assume (and later verify) that $f(E,\w)$ falls rapidly enough at large $\w$ that we can expand the exponent
in powers of $\w$,
\beq
S(E \pm \tfrac{1}{2} \w) = S(E) \pm \tfrac{1}{2} \Sp(E)\w+ \tfrac{1}{8} \Spp(E)\w^2 + \ldots,
\label{eq:SEpm}
\eeq
which yields
\beq
F_2(t)  = \frac{1}{Z}  \int_E e^{S(E)-\beta E} \! \int_{\w} e^{\Spp(E)\w^2\! /4} e^{i\w t}|f(E,\w)|^2.
\label{eq:F2bb}
\eeq
We do the $E$ integral by Laplace's method; 
this fixes $E$ to be the solution of $\Sp(E)=\beta$, which is the usual thermodynamic relation between energy and temperature. 
We can then also identify $\Spp(E)=-\beta^2\!/C$, where $C$ is the heat capacity of the system at inverse temperature $\beta$.
The remaining integral over $E$ yields a factor of the partition function $Z$. 
We therefore find
\beq
F_2(t)  = 
\int_\w e^{-\beta^2\w^2\! /4C}e^{i\w t}|f(E,\w)|^2. 
\label{eq:F2c}
\eeq 

Next we note that $\Tr(\rho A^2)$ is equal to $F_2(\pm\frac{i\beta}{2})$, and this should be a finite quantity. 
In the infinite-volume limit, $C\to\infty$, and we have
\beq
\Tr(\rho A^2) = \int_\w e^{\beta\w/2}|f(E,\w)|^2. 
\label{eq:TrA2}
\eeq 
For this to be finite, $f(E,\w)$ must fall at large $|\w|$ at least as fast as 
\beq
f(E,\w) \sim \exp(-\beta|\w|/4).
\label{eq:f2}
\eeq

We now consider the general four-point correlator for a single observable $A$
at inverse temperature $\beta$,
\beq
F_4(t_1,t_2,t_3) \coloneqq \Tr\big[ \rho^{1/4}\kern-2pt A(t_1) \rho^{1/4}\kern-2pt  A(t_2) \rho^{1/4}\kern-2pt  A(t_3) \rho^{1/4}\kern-2pt  A(0) \big] .
\label{eq:F4}
\eeq
Inserting four complete sets of energy eigenstates, we have
\beq
F_4 = \frac{1}{Z}\sum_{ijkl} e^{-\beta E} e^{i(\w_1t_1+\w_2t_2+\w_3t_3)}A_{ij}A_{jk}A_{kl}A_{li},
\label{eq:F4a} 
\eeq
where
$E \coloneqq \tfrac{1}{4} (E_i+E_j+E_k+E_l)$, 
$\w_1 \coloneqq E_i-E_j$, 
$\w_2 \coloneqq E_j-E_k$, 
$\w_3 \coloneqq E_k-E_l$.
We use Eq.~(\ref{eq:Aij}) with $\mathcal{A}(E)=0$ for $A_{ij}$. 
We then replace $R_{ij}R_{jk}R_{kl}R_{li}$ by its statistical average,
which, following the general analysis of Foini and Kurchan \cite{Foini2019}, we take to be
\beq
\overline{R_{ij}R_{jk}R_{kl}R_{li}} = \delta_{ik} + \delta_{jl} + e^{-S(E)}g(E, \w_1, \w_2, \w_3) .
\label{eq:RRRR}
\eeq
Here the first two terms account for the fact that for $i=k$ or $j=l$ the left-hand side reduces to the product of the absolute square of two $R$'s, 
and then the statistical average is 1.
The final term accounts for exponentially small correlations between different $R$'s.
Such correlations arise from treating the inner product of energy eigenstates (in a small range of energy)
with eigenstates of $A$ (in a small range of its eigenvalues) as a pseudorandom unitary matrix \cite{Foini2019}.
We view this as a generalized version of Berry's conjecture \cite{Berry1977} that the energy eigenstates of a chaotic quantum
system can be expressed as superpositions of suitable basis states whose coefficients are pseudorandom numbers with a gaussian distribution;
the specific form of the appropriate basis states is system dependent \cite{Hortikar1998a,Hortikar1998b,Gornyi2002}.
Berry's conjecture underlies the formulation of ETH presented in Ref.~\cite{Srednicki1994}, and so Eq.~(\ref{eq:RRRR})
should be viewed as a consequence of ETH.

Returning to Eq.~(\ref{eq:F4a}), we replace the sums by integrals, $\sum_i \to \int_{E_i} e^{S(E_i)}$,
expand the entropies to linear order about $E$, change
the integration variables to $E$ and the three $\w$'s, and perform the
integral over $E$ by Laplace's method. The final result, in the infinite volume limit, is
\begin{align}
F_4(t_1,t_2,t_3) &=
F_2(t_1 -t_2+ \tfrac{i \beta}{4})  F_2(t_3 + \tfrac{i \beta}{4}) 
\nonumber \\
&\quad + F_2(t_1 - \tfrac{i \beta}{4}) F_2(t_3-t_2- \tfrac{i \beta}{4})
\nonumber \\
&\quad+ F_{{\textsc{4C}}}(t_1,t_2,t_3),
\label{eq:F4c}
\end{align}
where the \emph{connected part} of the four-point function is
\begin{align}
F_{4\textsc{C}}(t_1,t_2,t_3) &= 
\int_{\w_1\cdots\w_3}  e^{i(\w_1t_1+\w_2t_2+\w_3t_3)} f(\w_1)f(\w_2)
\nonumber \\
&\hspace{4.5em} \times f(\w_3)f(-\w_1-\w_2-\w_3) 
\nonumber \\
&\hspace{4.5em} \times  g(\w_1,\w_2,\w_3).
\label{eq:F4C}
\end{align}
Here we have suppressed the $E$ dependence of $f$ and $g$.

Next we note that
\beq
\Tr(\rho A^4)-2[\Tr(\rho A^2)]^2 = F_{\textsc{4C}}(-\tfrac{i\beta}{4},-\tfrac{i\beta}{2},-\tfrac{3i\beta}{4})
\label{eq:TrA4}
\eeq
should be a finite quantity. Given Eq.~(\ref{eq:f2}), convergence of the integral over $\w_3$
in Eq.~(\ref{eq:F4C}), with $\w_1$ and $\w_2$ fixed, requires 
that $g(\w_1,\w_2,\w_3)$ must fall at large $|\w_3|$ at least as fast as 
\beq
g(\w_1,\w_2,\w_3) \sim \exp(-\beta|\w_3|/4).
\label{eq:g}
\eeq

We now turn our attention to the OTO four-point correlator
$F_{\textsc{oto}}(t) \coloneqq F_4(t,0,t)$, which is given by
\beq
F_{\textsc{oto}}(t) = 2 \mathop{\rm Re}[F_2(t+ \tfrac{i\beta}{4})^2] + F_{\textsc{4C}}(t,0,t).
\label{eq:FO3} 
\eeq
For times large compared to the dissipation time $t_\mathrm{d}$, which itself
should be comparable to or larger than $\beta$, the first term in Eq.~(\ref{eq:FO3}) will have decayed
to a negligible value, and we can replace $F_{\textsc{oto}}(t)$ with its connected part
${F}_{\textsc{otoc}}(t) \coloneqq F_{\textsc{4C}}(t,0,t)$.

We will be interested in the Fourier transform of ${F}_{\textsc{otoc}}(t)$, given by
\begin{align}
\widetilde{F}_{\textsc{otoc}}(\w) &\coloneqq
\int_{-\infty}^{+\infty}\frac{dt}{2\pi}\,e^{-i\w t}{F}_{\textsc{otoc}}(t)
\nonumber \\
&= \int_{\w_1,\,\w_2}f(\w_1)f(\w_2)f(\w-\w_1)f(-\w-\w_2)
\nonumber \\
&\hspace{4em} \times g(\w_1,\w_2,\w-\w_1).
\label{eq:tF}
\end{align}
From the large frequency behavior of $f$ and $g$ specified by Eqs.~(\ref{eq:f2})
and (\ref{eq:g}), we can infer that $\widetilde{F}_{\textsc{otoc}}(\w)$ must fall off at large $|\w|$
at least as fast as
\beq
\widetilde{F}_{\textsc{otoc}}(\w) \sim \exp(-3\beta|\w|/4).
\label{eq:tF2}
\eeq

To use this information, we need a more complete specification of the OTO correlator than is found in
Eq.~(\ref{eq:FO2}), which applies only for intermediate positive times. 
Assuming an exponential decay at late positive times, a simple model is 
$F_{\textsc{otoc}}(t) \propto 1/(1+z(t))^{\eta}$, where
\beq
z(t) \coloneqq e^{\lambda(t-t_\mathrm{s})},
\label{eq:z}
\eeq
and $\eta$ is a positive real parameter.
However this $F_{\textsc{otoc}}(t)$ is not time-reversal invariant, whereas Eq.~(\ref{eq:FO}) is.
To remedy this, and assuming $\lambda t_\mathrm{s}\gg 1$, we make the ansatz
\beq
F_{\textsc{otoc}}(t)={\cal N}G(z(t))G(z(-t)),
\label{eq:FGG}
\eeq
where we take $G(z)=1/(1+z)^{\eta}$; later we will consider other possibilities for $G(z)$.
The normalization constant is 
\beq
{\cal N}=\Tr\bigl[(\rho^{1/4}\!A){}^4\bigr]-2\bigl(\Tr\rho^{3/4}\!A\rho^{1/4}\!A\bigr){}^2. 
\label{eq:calN}
\eeq
From the product form of Eq.~(\ref{eq:FGG}), it follows that the Fourier transform is given by the convolution
\beq
\widetilde{F}_{\textsc{otoc}}(\w) = {\cal N}\int_{\w'} \widetilde{G}(\w-\w') \widetilde{G}(\w').
\label{eq:tFGG}
\eeq
From this it follows that the large-$\w$ behavior of $\widetilde{F}_{\textsc{otoc}}(\w)$ is the same as 
the large-$\w$ behavior of $\widetilde{G}(\w)$. We find
\beq
\widetilde{G}(\w) = e^{-i\w t_\mathrm{s}} \frac{ \Gamma(\eta+i\w/\lambda)\Gamma(0^+ -i\w/\lambda)}
{2\pi\lambda \, \Gamma(\eta)}K(\w),
\label{eq:tGw}
\eeq 
where $\Gamma(x)$ is the gamma function, and $K(\w)=1$ has been introduced for later convenience.
Eq.~(\ref{eq:tGw}) yields
$\widetilde{F}_{\textsc{otoc}}(\w) \sim \widetilde{G}(\w)\sim\exp(-\pi|\w|/\lambda)$ at large $|\w|$, independent of $\eta$. 
Requiring this fall-off to be at least as fast as Eq.~(\ref{eq:tF2}), we find the bound $\lambda\le 4\pi/3\beta$, 
which is more stringent than Eq.~(\ref{eq:lam}).
This shows that for $\lambda=2\pi/\beta$ (expected in conformal field theories with gravity duals),
the form $G(z)=1/(1+z)^\eta$, which has sometimes been used as an approximation (e.g., \cite{Roberts2014}),
is inconsistent with our Eq.~(\ref{eq:tF2}).

Maldacena \textit{et al.}~\cite{Maldacena2016b}
computed the OTO correlator for a conformal field with dimension $\Delta=\eta/2$ via the 
$\mathrm{AdS}_2$ gravity dual. In our notation, their result is
\beq
G(z) = \int_0^\infty du\,h(u)(1+ u z)^{-\eta}
\label{eq:G}
\eeq
with $h(u)= e^{-u}u^{\eta-1}\!/\Gamma(\eta)$. This yields Eq.~(\ref{eq:tGw}) with
\beq
K(\w) = \int_0^\infty du\,h(u)u^{i\w/\lambda}.
\label{eq:K}
\eeq
For the $h(u)$ of Ref.~\cite{Maldacena2016b},
$K(\w)=\Gamma(\eta+i\w/\lambda)/\Gamma(\eta)\sim \exp(-\pi|\w|/2\lambda)$,
and hence $\widetilde{F}_{\textsc{otoc}}(\w) \sim\exp(-3\pi|\w|/2\lambda)$.
Requiring this fall-off to be at least as fast as Eq.~(\ref{eq:tF2}), we find the bound 
$\lambda\le 2\pi/\beta$, the same as Eq.~(\ref{eq:lam}).

More generally, the bound $\lambda\le 2\pi/\beta$ holds if 
$K(\w)\sim \exp(-c\pi|\w|/2\lambda)$ with $c\le 1$ at large $\w$.
The Paley--Wiener theorem \cite{Reed1975} implies that this will be the case
if and only if there is a value of $\theta\in[-\pi/2,\pi/2]$ such that
\beq
\int_0^\infty du\,u|h(e^{i\theta}u)|^2=\infty.
\label{eq:inth}
\eeq
For example, this is the case if $h(u)\sim u^a\exp(-b u^\gamma)$ at large $u$ with 
$a\ge -\frac12$, $b\ge 0$ and $\gamma\ge 1$.
However, we have not been able to connect the mathematical condition of Eq.~(\ref{eq:inth}) 
to a physical property of the system.

We can compare our derivation of Eq.~(\ref{eq:lam}) with that of Ref.~\cite{Maldacena2016}.
A key assumption used in Ref.~\cite{Maldacena2016}, their Eq.~(23), is an approximate factorization of
a different regularization of the four-point function at intermediate times $t_\mathrm{d} \ll t \ll t_\mathrm{s}$,
\beq
\Tr\!\big[ \rho^{1/2}\kern-2pt A(t) A(0) \rho^{1/2}\kern-2pt  A(0) A(t) \big] 
\le \bigl(\Tr\rho^{1/2}\kern-2pt A \rho^{1/2}\kern-2pt  A\bigr){}^2 +\varepsilon,
\label{eq:23}
\eeq
where $\varepsilon$ is a small tolerance parameter. In our notation, this is equivalent to
\beq
|F_2(t- \tfrac{i \beta}{2})|^2 + F_{{\textsc{4C}}}(t -\tfrac{i \beta}{4},t, -\tfrac{i \beta}{4}) \le \varepsilon.
\label{eq:23a}
\eeq
We expect the first term to be negligible at the relevant intermediate times, and it is plausible that the second term,
which is an in-time-order connected four-point function, is also negligible.
However, we do not need to assume this in our analysis.

Assuming Eq.~(\ref{eq:23}), Ref.~\cite{Maldacena2016} then establishes that 
the rescaled correlator $f(t)\coloneqq F_\textsc{oto}(t)/[F_2(0)^2+\varepsilon]$ obeys
\beq
\bigg|\frac{d}{dt} f(t)\bigg| \le \frac{2\pi}{\beta}\coth\!\bigg(\frac{2\pi t}{\beta}\bigg)\frac{1-f(t)^2}{2}.
\label{eq:dfdt}
\eeq
Because of the factor of $1-f^2$ on the right-hand side of Eq.~(\ref{eq:dfdt}), 
it provides a meaningful bound on $\lambda$ in Eq.~(\ref{eq:FO2}) if and only if
$f(t)=1 - e^{\lambda (t-t_\mathrm{s})}$ for $t_\mathrm{d}\ll t\ll t_\mathrm{s}$,
where we now have an equality rather than a proportionality.
Equivalently, $F_\textsc{oto}(t)$ must be very close to $F_2(0)^2$ for these intermediate times.
In our notation, this requires ${\cal N} \approx F_2(0)^2$, where $\cal N$ is the normalization constant of Eq.~(\ref{eq:calN}).
Such a relation would follow from large-$N$ factorization, Eq.~(9) in Ref.~\cite{Maldacena2016},
but is not expected to be true more generally. 
Our derivation of Eq.~(\ref{eq:lam}) does not require this assumption.

We note that the scrambling time $t_\mathrm{s}$ appears as the period of an oscillation in the
amplitude of $\widetilde{F}_{\textsc{otoc}}(\w)$, cf.~Eqs.~(\ref{eq:tFGG},\ref{eq:tGw}), 
that must have its origin in a correspoding oscillation in the amplitude of $g(\w_1,\w_2,\w_3)$, 
cf.~Eq.~(\ref{eq:tF}). The underlying physics of this sort of oscillation in the four-point correlation of 
operator matrix elements, cf.~Eq.~(\ref{eq:RRRR}), is worthy of further exploration.

Recently another measure of chaos was introduced by Parker \textit{et al.}~\cite{Parker2019}:
an operator complexity growth rate $\alpha$ that was shown to be bounded by $\alpha \le \pi/\beta$.
This bound on $\alpha$ is related to the the large-$\w$ behavior of $f(E,\w)$, and follows from Eq.~(\ref{eq:f2}). 
Parker \textit{et al.}~conjecture that, very generally, $\lambda \leq 2\alpha$.
Our analysis shows that the bound on $\lambda$ requires information
about the large-$\w$ behavior of $g(E,\w_1,\w_2,\w_3)$ in addition to the large-$\w$ behavior of $f(E,\w)$.
A proof that $\lambda \leq 2\alpha$ would therefore imply a further relationship between these two functions,
and hence further structure in the ETH matrix elements.
This is an interesting topic for further research.

To conclude, we have derived the known bound of Eq.~(\ref{eq:lam}) 
on the growth rate of the out-of-time-order four-point correlator
from the structure of operator matrix elements that follows from the
eigenstate thermalization hypothesis, Eqs.~(\ref{eq:Aij},\ref{eq:RRRR}).
We also needed a mild assumption on the functional form of this correlator, 
as specified by Eqs.~(\ref{eq:FGG},\ref{eq:G},\ref{eq:inth}).
However, we did not need the assumptions of Eq.~(\ref{eq:23}) and of large-$N$ factorization (or its equivalent) 
that were needed in the analysis of Ref.~\cite{Maldacena2016}.

We hope that this unification of two key paradigms of thermal behavior in many-body quantum systems
will lead to further insights into this important branch of physics.

\begin{acknowledgments}

We thank Xiangyu Cao, Sergio Hernandez-Cuenca, Jorge Kurchan, Silvia Pappalardi, Daniel E.~Parker, Eduardo Test\'e,
and Gustavo J.~Turiaci for helpful discussions.
This work was supported in part by the Microsoft Corporation Station Q (C.M.).

\end{acknowledgments}

\end{document}